# On A Saturated Poromechanical Framework and Its Relation to Abaqus Soil Mechanics and Biot Poroelasticity Frameworks

Lei Jin[1]

[1]ExxonMobil Upstream Research Company, Spring, TX 77389, USA

lei.jin@exxonmobil.com

**Abstract**

We introduce a conservational and constitutive framework for a closed and isothermal two-phase material system consisting of a deformable porous solid matrix and a fully saturating, single-phase, and compressible pore fluid without inter-phase mass exchange. We re-derive a generalized fluid mass balance law using fundamental transport rules. We also summarize from the literature a fundamental force balance law for the fluid-solid mixture that does not require any effective stress law *a priori*. We show that the two conservation laws are coupled naturally to second-order without any constitutive prerequisites. This differs from Biot poroelasticity, which first postulates first-order fluid-solid coupling as two linearized constitutive relationships and then enforces them into simple Eulerian form of conservation laws. Next, we examine a limiting-case unsaturated soil mechanics framework implemented in Abaqus, by assuming isothermal conditions, full saturation, and no adsorption, and then relate it to our framework. We prove that (1) the two mass balance laws are always equivalent regardless of fluid constitutive behaviors, and (2) the two force balance laws are equivalent in their specific forms with a linearly elastic solid skeleton. Finally, taking advantage of a fundamental pore constitutive law, we show how our framework, and by extension the limiting-case Abaqus framework, naturally gives rise to the distinction between drained and undrained settings, and reduces to Biot poroelasticity under simplifying conditions. Notably, our framework indicates the presence of an additional solid-to-fluid coupling term when the solid particle velocity is non-orthogonal to the Darcy velocity.

**Keywords**: poromechanics, soil mechanics, Biot poroelasticity, fluid-solid coupling, Abaqus

# 1. Introduction

Poromechanics describes the kinematic and deformational behavior of a porous solid with interconnected pores that are either partially or fully saturated with a single- or multi-phase fluid. Poromechanics is critical to the study of geomaterials such as soils and rocks, engineering materials like concretes and ceramics, as well as various biological tissues including cartilage and cornea. De Boer (1996) reviewed in part some highlights in the historical developments of poromechanical theories for saturated porous media under infinitesimal deformation. Critical milestones include Terzaghi's soil mechanics framework (Terzaghi, 1943), and Biot's poroelasticity (Biot, 1941; 1955) and its subsequent recasting (Rice & Cleary,1976). More recently, Coussy (2004) developed a thermodynamically consistent poromechanical framework, by treating the solid porous continuum as an open system whose energy balance and entropy imbalance are perturbed by the addition of fluid phases. This framework offers a natural extension to partial saturation and finite deformation ranges.

An essential element of a poromechanical governing framework is the hydro-mechanical interaction (i.e., coupling) between the fluid and solid constituents. Biot (1941) seminally postulated two linearized constitutive relationships to describe this coupling effect. Specifically, fluid content depends linearly on fluid pressure and mean solid stress, and solid strain depends linearly on solid stress and fluid pressure. The constitutive pair are retained in Rice & Cleary (1976). Upon substitution into simple Eulerian forms of conservation laws, they suggest that the negative pore pressure gradient, which points from high pressure to low pressure, acts as an equivalent body force to drive deformation, strain, and stress, whereas a fraction of the volumetric strain rate acts as an equivalent fluid source or sink (Wang, 2000; Segall, 2010). One critical question is whether the coupling arises naturally from the conservation laws, irrespective of the constituents and their behaviors. Whether this coupling obeys the linear forms postulated by Biot, also remains an interesting open question.

Recently, there has been some renewed interest in Biot poroelasticity in the geophysics community, driven by the need for mechanistic studies of fluid injection-related seismicity (e.g., Ellsworth, 2013). Segall & Lu (2015) first demonstrated the poroelastic stressing effect as an important triggering mechanism of such seismicity. They used the classic Rudnicki (1986) analytical solution to Biot's governing equations, which is derived for a point source of fluid mass rate embedded at the center of an isotropic, homogeneous full space that is linearly elastic for the solid matrix and linearly diffusive for the pore fluid. Later studies implemented Biot poroelasticity accounting for physical realism through numerical modeling. A handful of useful numerical solvers have been developed for Biot poroelasticity with sophisticated model configurations (e.g., Wang & Kümpel, 2003; Jin & Zoback, 2017). Some other studies opted to implement Biot poroelasticity using Abaqus (e.g., Fan et. al., 2016; Haddad & Eichhubl, 2020), a commercial finite element solver (Dassault Systèmes Simulia, 2014), based on studies claiming that Abaqus can reproduce classic solutions to Biot poroelasticity with appropriate parameter tuning, i.e., mapping Biot's parameters to equivalent Abaqus parameters (Altmann, 2010; Altmann et al., 2010).

Such a practice reveals some apparent confusion among poromechanics practitioners regarding the underlying governing laws Abaqus solves. In fact, Abaqus does not implement linear Biot poroelasticity but instead, a general nonlinear unsaturated soil mechanics framework formulated for a three-phase mixture consisting of solid grains, a wetting fluid, and air. The fluid-solid coupling is also two-way but is radically different from Biot's postulations. First, the overall material deforms in response to the saturation-dependent effective stress (Wu, 1967), dissimilar to the Biot effective stress (Biot, 1941; Nur & Byerlee 1971; Berryman, 1992); second, the solid grains respond to local fluid pressure directly via intrinsic density changes, in contrast to Biot's solid strain - fluid pressure relationship. Two additional distinctions are also worth noting. First, the effective stress in Abaqus drives changes not in strain, as is the case in Biot poroelasticity, but the so-called *effective strain*; second, Abaqus also considers behaviors such as fluid entrapment and thermal expansion of the fluid and solid grains (Dassault Systèmes Simulia, 2014).

For these reasons, there appears to be a need for a general poromechanical conservational framework free from constitutive postulations. Here, we first offer some insights into this issue. Also, there appears to be a gap to be bridged among a general framework, Biot's framework, and the Abaqus framework. Since Biot poroelasticity assumes a single-phase fluid, full saturation, and infinitesimal deformation, we shall restrict our focus as such. For meaningful comparisons, we obtain a limiting case of the Abaqus framework by considering full saturation and no fluid entrapment nor thermal expansion. We then provide a mathematical description of how these three frameworks relate to one another. The goal of this study is to help practitioners from different communities understand the common underlying problem, promote clarity, and facilitate the appropriate use of commercial tools.

## 2. A Saturated Poromechanical Framework

### 2.1 Basic definitions and relationships

Consider a solid porous medium in which the pore space amongst constituent solid grains is fully saturated with a single-phase fluid. The fluid moves freely relative to the solid without trapping or adsorption. The material is an idealized two-phase (fluid, solid) mixture. Three reference frames can be defined, including two Lagrangian reference frames following the motion of fluid and solid particles, respectively, and a Eulerian reference frame. Consider a material point $X_j$ ($j=f, s$ indicates the phase) with a position vector $X_j$ in a phase-$j$ Lagrangian reference frame, and denote its current position vector in a Eulerian reference frame as $x$, such that

$$x = \boldsymbol{\varphi}_j(X_j), \quad X_j = \boldsymbol{\varphi}_j^{-1}(x), \quad j = f, s \tag{1}$$

where $\boldsymbol{\varphi}_j$ is a unique and invertible vectorial mapping function.

The phase-$j$ particle velocity is the same in either reference frame. This reads

$$V_j(\boldsymbol{X}_j,t) = \frac{\partial \boldsymbol{\varphi}_j(\boldsymbol{X}_j,t)}{\partial t} = \boldsymbol{v}_j(\boldsymbol{x},t) = \frac{\partial \boldsymbol{\varphi}_j\left(\boldsymbol{\varphi}_j^{-1}(\boldsymbol{x}),t\right)}{\partial t}, \quad j=f,s \tag{2}$$

where $\boldsymbol{V}_j$ is the phase-$j$ Lagrangian velocity, and the $\boldsymbol{v}_j$ is the associated Eulerian velocity. Since the two are conveniently equal, we use the latter by default subsequently. When fluid particles travel in pores, $\boldsymbol{v}_f$ is also referred to as the *interstitial velocity*.

Definitions of a gradient operator, $\nabla$, and a divergence operator, $\nabla \cdot$, also depend on the reference frame. For brevity, these two operators throughout this study always default back to their definitions in the Eulerian reference frame, which reads

$$\begin{aligned}\nabla &:= \nabla_x = \frac{\partial}{\partial \boldsymbol{x}} \\ \nabla \cdot &:= \nabla_x \cdot = \frac{\partial}{\partial \boldsymbol{x}} \cdot\end{aligned} \tag{3}$$

The deformation gradient $\boldsymbol{F}$ and the associated Jacobian $J$ are defined by default only for the solid phase as

$$\boldsymbol{F} := \boldsymbol{F}_s = \frac{\partial \boldsymbol{x}}{\partial \boldsymbol{X}_s} = \nabla_{X_s}\boldsymbol{x} \tag{4}$$

$$J = \det(\boldsymbol{F}) \tag{5}$$

The material time derivative (i.e., particle derivative, total time derivative, or substantial derivative) in a two-phase material system is defined differently, depending on the phase. It is related to the Eulerian partial time derivative as

$$\frac{d^j}{dt}(\square) = \frac{\partial}{\partial t}(\square) + \boldsymbol{v}_j(\boldsymbol{x},t) \cdot \nabla(\square), \quad j=f,s \tag{6}$$

Writing out equation (6) for both phases and subtracting the two leads to the following, which will soon prove useful

$$\frac{d^f}{dt}(\square) - \frac{d^s}{dt}(\square) = \left(\boldsymbol{v}_f(\boldsymbol{x},t) - \boldsymbol{v}_s(\boldsymbol{x},t)\right) \cdot \nabla(\square) \tag{7}$$

In this study, the default total time derivative is defined by tracking the motion of solid particles as

$$\dot{\square} = \frac{d}{dt}(\square) := \frac{d^s}{dt}(\square) \tag{8}$$

For the solid phase, the following relationship between the material time derivative of the Jacobian and the Eulerian velocity is well established (e.g., Hughes, 2012)

$$\dot{J} = \frac{dJ}{dt} = \frac{d^s J}{dt} = J\nabla \cdot \boldsymbol{v}_s(\boldsymbol{x},t) \tag{9}$$

**2.2 Mass conservation for the fluid phase**

Consider in the current Eulerian reference frame an arbitrary material volume of $V(\boldsymbol{x}, t) = V_f(\boldsymbol{x}, t) + V_s(\boldsymbol{x}, t)$, where $V_f(\boldsymbol{x}, t)$ and $V_s(\boldsymbol{x}, t)$ are the fluid and solid volumes, respectively. Because the pore space is fully saturated, the fluid volume $V_f(\boldsymbol{x}, t)$ is identical to the pore volume $V_\phi(\boldsymbol{x}, t)$. We define

$$\phi(\boldsymbol{x},t) := \frac{V_\phi(\boldsymbol{x},t)}{V(\boldsymbol{x},t)} = \phi_f(\boldsymbol{x},t) := \frac{V_f(\boldsymbol{x},t)}{V(\boldsymbol{x},t)} \tag{10}$$

where $\phi(\boldsymbol{x}, t)$ is the Eulerian porosity and $\phi_f(\boldsymbol{x}, t)$ is the current fluid volumetric fraction.

We now make three assumptions: (1) the system is isothermal, and materials are barotropic, (2) no interphase mass exchange occurs, and (3) the system is closed and free from external mass sources. Under these, by following the motion of solid particles and considering equation (10), the conservation of fluid mass within the material volume, denoted as $m_f$, reads

$$0 = \dot{m}_f = \frac{d^s m_f}{dt} = \frac{d^s}{dt} \int_{V(\boldsymbol{x},t)} \rho_f(\boldsymbol{x},t) \phi_f(\boldsymbol{x},t) dV \tag{11}$$

where $\rho_f$ is the intrinsic density of the fluid.

In the same Eulerian reference frame, consider also an arbitrary hosting porous solid skeleton control volume $CV(\boldsymbol{x}, t)$ with a control surface $CS(\boldsymbol{x}, t)$. Note that $CV(\boldsymbol{x}, t)$ is not necessarily the same as the material volume $V(\boldsymbol{x}, t)$. Therefore, equation (11) can be expanded according to the *general Reynolds transport theorem* as

$$0 = \frac{d^s}{dt} \int_{CV(\boldsymbol{x},t)} \rho_f(\boldsymbol{x},t) \phi_f(\boldsymbol{x},t) dV + \int_{CS(\boldsymbol{x},t)} \rho_f(\boldsymbol{x},t) \phi_f(\boldsymbol{x},t) \tilde{\boldsymbol{v}}(\boldsymbol{x},t) \cdot \boldsymbol{n} dS \tag{12}$$

where $\boldsymbol{n}$ is the unit normal vector and $\tilde{\boldsymbol{v}}(\boldsymbol{x},t)$ is the fluid particle velocity relative to that of the moving solid control surface, that is

$$\tilde{\boldsymbol{v}}(\boldsymbol{x},t) := \boldsymbol{v}_f(\boldsymbol{x},t) - \boldsymbol{v}_s(\boldsymbol{x},t) \tag{13}$$

Here, $\tilde{\boldsymbol{v}}$ is also referred to as the *seepage velocity*.

Meanwhile, the first term on the right-hand side of (12) requires differentiation under the integral sign, which can be carried out according to the *Leibniz theorem* as

$$\frac{d^s}{dt}\int_{CV(\boldsymbol{x},t)}\rho_f(\boldsymbol{x},t)\phi_f(\boldsymbol{x},t)dV$$
$$=\int_{CV(\boldsymbol{x},t)}\left[\frac{\partial}{\partial t}\left(\rho_f(\boldsymbol{x},t)\phi_f(\boldsymbol{x},t)\right)\right]dV+\int_{CS(\boldsymbol{x},t)}\left[\rho_f(\boldsymbol{x},t)\phi_f(\boldsymbol{x},t)v_s(\boldsymbol{x},t)\cdot\boldsymbol{n}\right]dS \quad (14)$$

The surface integrals in equations (12) and (14) can be converted to volume integrals according to the *divergence theorem*, therefore, equation (11) now becomes

$$0=\int_{CV(\boldsymbol{x},t)}\left[\frac{\partial}{\partial t}\left(\rho_f(\boldsymbol{x},t)\phi_f(\boldsymbol{x},t)\right)+\nabla\cdot\left(\rho_f(\boldsymbol{x},t)\phi_f(\boldsymbol{x},t)\boldsymbol{v}_s(\boldsymbol{x},t)\right)+\nabla\cdot\left(\rho_f(\boldsymbol{x},t)\phi_f(\boldsymbol{x},t)\tilde{\boldsymbol{v}}(\boldsymbol{x},t)\right)\right]dV \quad (15)$$

Since equation (15) holds true for any arbitrary $CV(\boldsymbol{x},t)$, we have the following Eulerian form of fluid mass balance (hereinafter, we omit writing $(\boldsymbol{x},t)$ for brevity)

$$\frac{\partial}{\partial t}\left(\rho_f\phi_f\right)+\nabla\cdot\left(\rho_f\phi_f\boldsymbol{v}_s\right)+\nabla\cdot\left(\rho_f\phi_f\tilde{\boldsymbol{v}}\right)=0 \quad (16)$$

Equation (16) can be also written in an equivalent arbitrary solid-Lagrangian Eulerian form ($AL_sE$) as

$$\frac{d^s}{dt}\left(\rho_f\phi_f\right)+\left(\rho_f\phi_f\right)\nabla\cdot\boldsymbol{v}_s+\nabla\cdot\left(\rho_f\phi_f\tilde{\boldsymbol{v}}\right)=0 \quad (17)$$

Alternatively, and perhaps more intuitively, equation (17) can be derived by tracking the motion of fluid particles themselves. The fluid mass conservation now reads

$$0=\dot{m}_f=\frac{d^f m_f}{dt}=\frac{d^f}{dt}\int_V \rho_f\phi_f dV$$
$$=\int_V\left[\frac{\partial}{\partial t}\left(\rho_f\phi_f\right)\right]dV+\int_S\left[\rho_f\phi_f\boldsymbol{v}_f\cdot\boldsymbol{n}\right]dS \quad (18)$$
$$=\int_V\left[\frac{\partial}{\partial t}\left(\rho_f\phi_f\right)+\nabla\cdot\left(\rho_f\phi_f\boldsymbol{v}_f\right)\right]dV$$

Here, the fluid material derivative of the volume integral is expanded directly following the Leibniz theorem, followed again by applying the divergence theorem. The arbitrariness of the material volume $V$ in equation (18) mandates that

$$\frac{\partial}{\partial t}\left(\rho_f\phi_f\right)+\nabla\cdot\left(\rho_f\phi_f\boldsymbol{v}_f\right)=0 \quad (19)$$

Equation (19) can be re-written in an arbitrary fluid-Lagrangian Eulerian form ($AL_fE$) as

$$\frac{d^f}{dt}(\rho_f \phi_f) + (\rho_f \phi_f) \nabla \cdot \mathbf{v}_f = 0 \tag{20}$$

Based on equations (7) and (13), equation (20) can be expanded as

$$\frac{d^s}{dt}(\rho_f \phi_f) + \tilde{\mathbf{v}} \cdot \nabla (\rho_f \phi_f) + (\rho_f \phi_f) \nabla \cdot (\mathbf{v}_s + \tilde{\mathbf{v}}) = 0 \tag{21}$$

Equation (21) can be trivially proven equivalent to equation (17), and hence, equation (16).

Finally, in the case of external fluid mass perturbation, the third assumption on the closed material system can be relaxed by replacing the 0 on the *RHS* of equations (16), (17), and (19) ~ (21) with a fluid mass rate term.

**2.3 Force equilibrium for the continuum of fluid-solid mixture**

A general form of the linear momentum balance law has been rigorously derived for an unsaturated fluid-solid mixture without any prior assumption (Borja, 2006). In the limits of full saturation and quasi-static particle motion (i.e., inertia ≈ 0), it reduces to the following force balance law in a Eulerian reference frame (again, ($\mathbf{x}$, $t$) is omitted)

$$\nabla \cdot \boldsymbol{\sigma} + \underbrace{(\rho_f \phi_f + \rho_s (1 - \phi_f))}_{\rho_{mix}} \mathbf{g} \simeq \mathbf{0} \tag{22}$$

where $\boldsymbol{\sigma}$ is the 2nd-order total Cauchy stress tensor, $\mathbf{g}$ is the gravitational acceleration, $\rho_f$ is the same as before, $\rho_s$ is the intrinsic density of solid grains, and $\rho_{mix}$ is the density of the fluid-solid mixture as a summation of the so-called partial densities $\rho^f = \rho_f \phi_f$ and $\rho^s = \rho_s (1 - \phi_f)$.

The key then lies in the formulation of an effective stress law that describes how total stresses are distributed to solid grains in the presence of fluid pressure. Several seminal studies exist. The first rigorous derivation was given by Nur & Byerlee (1971) in which the inherent phenomenological assumption of linear dependence of strain on stress and pore pressure made by Biot (1941) was retained (details in section 5.1). A more rigorous derivation without such an assumption was arrived from the first and second laws of thermodynamics by Coussy (2004; 2007) and Borja (2006), respectively. For a fully saturated porous skeleton, these studies arrived at the same effective stress law. Following a *tension positive* notation, it reads

$$\boldsymbol{\sigma}' = \boldsymbol{\sigma} + \underbrace{\left(1 - \frac{K}{K_s}\right)}_{\alpha} p_f \mathbf{1} \tag{23}$$

where $\boldsymbol{\sigma}'$ is the 2nd-order effective Cauchy stress tensor, $p_f$ is the pore fluid pressure, **1** is the second-order unit identity, $K_g$ is the bulk modulus of solid grains, and $K$ is the bulk modulus of the porous solid skeleton (i.e., matrix), sometimes referred to as the dry bulk modulus or the drained bulk modulus. $1-K/K_s$ is the bulk volumetric effective stress coefficient, which coincides with the widely known form of the Biot-Willis coefficient $\alpha$ (Biot & Willis, 1957), see also section 5.1.

It was shown that $\boldsymbol{\sigma}'$ arises as a power-conjugate to the solid's rate of deformation tensor $d=(\nabla \boldsymbol{v}_s + \boldsymbol{v}_s \nabla)/2$. Substituting equation (23) into equation (22) and contracting the divergence operator with the unit identity lead to

$$\nabla \cdot \boldsymbol{\sigma}' - \left(1 - \frac{K}{K_s}\right)\nabla p_f + \left(\rho_f \phi_f + \rho_s (1-\phi_f)\right)\mathbf{g} = \mathbf{0} \qquad (24)$$

## 2.4 The natural arising of 1st- and 2nd-order fluid-solid full coupling

Equation (17) (or its equivalent forms by equations (16), (19), (20)) and equation (24) are the two strong forms of conservation laws governing a general, single-phase, saturated, transient, and quasi-static poromechanical problem under isothermal conditions and without inter-phase mass exchanges nor external fluid mass perturbations. Both equations are to be closed with appropriate Dirichlet and Neumann boundary conditions. The presence of their interacting terms shows that the fluid-solid full coupling arises naturally without any prerequisite. This is fundamentally different from the approach in Biot (1941), Rice & Cleary (1976), and Cleary (1977), where coupling terms in the two conservation laws are "imposed" through direct substitution of two linearized phenomenological constitutive laws (i.e., fluid content depends linearly on fluid pressure and mean solid stress, and solid strain depends linearly on solid stress and fluid pressure, see also section 5.1).

Furthermore, the two equations here reveal two orders of full coupling, as opposed to one order of coupling in Biot poroelasticity. To the 1st-order, the divergence of solid particle velocity, $\nabla \cdot \boldsymbol{v}_s$, which represents the rate of volumetric change in the hosting porous solid skeleton, when scaled by $\rho_f \phi_f$, acts as an equivalent fluid source in the form of mass per unit volume per unit time. This solid-to-fluid coupling fundamentally arises due to the motion (including both rigid translation and deformation) of the hosting porous solid. On the other hand, the pore pressure gradient, when scaled by $1-K/K_s$, acts as an equivalent body force (force per unit volume). The 2nd-oder coupling, albeit more subtle, arises from $\phi_f$ appearing in both equations. $\phi_f$ is an inherent function of $p_f$ and $\boldsymbol{v}_s$. For example, expanding the first term in equation (17) requires a tangent constitutive law for $d\phi_f$ (see section 2.5), which has been proven by Coussy (2004) under certain conditions as a linear combination of $dp_f$ and $d\varepsilon_{\text{vol}}$ ($\varepsilon_{\text{vol}} = tr(\boldsymbol{\varepsilon})$ is the volumetric strain) and $d\varepsilon_{\text{vol}}/dt$ relates to $\nabla \cdot \boldsymbol{v}_s$, thereby introducing a second equivalent fluid source in the fluid mass balance due to solid motion. Meanwhile, the body force in equation (24) depends on the fluid because the mixture density evolves with pore pressure changes. One way to clearly see this coupling effect is by examining changes in the associated gravitational force vector $\mathbf{F}_g$ acting over an arbitrary chunk of current material volume

$$d\mathbf{F}_g = d\int_{V(x,t)} \rho_{mix}(\boldsymbol{x},t)\mathbf{g}dV \tag{25}$$

For differentiation of the volume integral, we perform volume transport back and forth between Lagrangian and Eulerian configurations. This reads (here, the reference frame is written out here to show the pull-back and push-forward operations, and subscript "0" indicates a previous time step)

$$\begin{aligned}d\mathbf{F}_g &= d\int_{V_0(X,t)} \rho_{mix}(\boldsymbol{X},t)\mathbf{g}JdV_0 \\ &= \int_{V_0(X,t)} d\left(\rho_{mix}(\boldsymbol{X},t)J\right)\mathbf{g}dV_0 \\ &= \int_{V_0(X,t)} \left[J_0 d\left(\rho_{mix}(\boldsymbol{X},t)\right) + \rho_{mix0}(\boldsymbol{X},t)dJ\right]\mathbf{g}dV_0 \\ &= \int_{V(x,t)} \left[Jd\left(\rho_{mix}(\boldsymbol{x},t)\right) + \rho_{mix0}(\boldsymbol{x},t)dJ\right]\frac{1}{J}\mathbf{g}dV\end{aligned} \tag{26}$$

Substituting equation (9) into (26) leads to

$$d\mathbf{F}_g = \int_V \left[d\rho_{mix} + \rho_{mix0}\nabla\cdot\boldsymbol{v}_s dt\right]\mathbf{g}dV \tag{27}$$

Here, the 2$^{nd}$-order fluid-to-solid coupling is introduced by $d\rho_{mix}$, which can be shown trivially as

$$d\rho_{mix} = (\rho_{f0} - \rho_{s0})d\phi_f + \phi_{f0}d\rho_f + (1-\phi_{f0})d\rho_s \tag{28}$$

Equation (28) shows three specific sources of this coupling effect, including $d\phi_f$ that relates to $dp_p$ through a tangent constitutive law, and $d\rho_f$, $d\rho_s$ that relate to $dp_f$ via material compressibilities, as are detailed subsequently in equations (42) and (43).

## 2.5 Constitutive laws

Equations (17) and (24) can be solved for two primary unknowns of interest related to fluid and solid phases, respectively. For the fluid, the obvious choice is $p_p$. For the solid, the choice can vary, and one convenient option is the displacement vector, which is defined as

$$\boldsymbol{u} := \boldsymbol{x} - \boldsymbol{X}_s = \boldsymbol{\varphi}_s(\boldsymbol{X}_s) - \boldsymbol{X}_s \tag{29}$$

Both conservation laws take more specific forms upon substitutions of appropriate fluid, solid, and pore constitutive laws. In equation (17), the product of the differential Eulerian velocity $\tilde{v}$ and the Eulerian porosity $\phi_f$ is known as the *Darcy velocity*, denoted here as $\widetilde{\underline{v}}$, which relates to $p_f$ as

$$\widetilde{\underline{v}} := \phi_f \tilde{v} = -\eta^{-1}\boldsymbol{k}\cdot(\nabla p_f + \rho_f \mathbf{g}) \tag{30}$$

where $\eta$ is the fluid viscosity, and $\boldsymbol{k}$ is the 2nd-order permeability tensor permitted to be fully anisotropic, simply anisotropic (with only diagonal elements), or isotropic, in which case $\boldsymbol{k}=k\mathbf{1}$.

Equation (30) states a linear fluid constitutive law. On the other hand, if we assume the solid skeleton to be linearly elastic and isotropic, then the Hooke's law can be chosen for describing its constitutive behavior. It reads

$$\boldsymbol{\sigma}' = \mathbb{C}^e : \boldsymbol{\varepsilon} = \left[ K\mathbf{1}\otimes\mathbf{1} + 2\mu\left(\mathbf{I} - \frac{1}{3}\mathbf{1}\otimes\mathbf{1}\right) \right] : \boldsymbol{\varepsilon} \tag{31}$$

where $K$ is same as before, $\mu$ is the shear modulus of the porous solid skeleton (i.e., dry or drained shear modulus), $\mathbf{I}$ and $\mathbf{1}$ are 4th-order and 2nd-order unit identities, respectively, $\mathbb{C}^e$ is the 4th-order elastic stiffness tensor decomposed into isotropic and deviatoric parts, and $\boldsymbol{\varepsilon}$ is the 2nd-order Eulerian strain tensor.

If we further assume infinitesimal deformation, then two simplifications can be made. First, $\boldsymbol{\varepsilon}$ can be approximated by ignoring 2nd-order terms as the symmetric gradient of the displacement

$$\boldsymbol{\varepsilon} \approx \nabla^{(s)}\boldsymbol{u} = \left[\frac{1}{2}\left(\nabla\boldsymbol{u} + (\nabla\boldsymbol{u})^T\right)\right] \tag{32}$$

Second,

$$\partial\boldsymbol{u} \simeq \partial\boldsymbol{\varphi}_s,\ \boldsymbol{v}_s \simeq \frac{\partial\boldsymbol{u}}{\partial t},\ \nabla\cdot\boldsymbol{v}_s \simeq \nabla\cdot\left(\frac{\partial\boldsymbol{u}}{\partial t}\right) = \frac{\partial(\nabla\cdot\boldsymbol{u})}{\partial t} = \frac{\partial\varepsilon_{vol}}{\partial t} \tag{33}$$

By now, the poromechanical system becomes linearly poroelastic. Material linearity is given by equations (30) and (31) whereas geometric linearity is given by equations (32) and (33).

As mentioned in section 2.4, an incremental pore constitutive law is required to carry out the differentiation. From the second law of thermodynamics, Coussy (2004) proved the existence of the following linear relationship when assuming no energy dissipation, isothermal conditions, and infinitesimal deformation and isotropy of the solid skeleton. It reads

$$d^s\phi_f = \frac{\alpha - \phi_{L0}}{K_s}d^s p_f + (\alpha - \phi_f)d^s\varepsilon_{vol} \tag{34}$$

Here, $\phi_L$ is the so-called *Lagrangian porosity*, defined as the current pore volume divided by the initial total volume, and can be related to the Eulerian porosity through the following transport. Carried out for the initial Lagrangian porosity $\phi_{L0}$, it reads

$$\phi_{L0} = J_0\phi_f = \det(\mathbf{1})\phi_f = \phi_f \tag{35}$$

Jin, Poromechanical FrameworksSubstituting equation (35) into equation (34) recovers the commonly seen constitutive law for $d\phi_f$ as a linear combination of $dp_f$ and $d\boldsymbol{\varepsilon}_{vol}$ (e.g., Wang, 2000; Cheng, 2016; Zhang et al., 2021), see equation (62).

## 3. Abaqus Soil Mechanics: The Limiting Case

Abaqus offers the capacity to model a general three-phase unsaturated soil mechanics problem (Dassault Systèmes Simulia, 2014). The medium of interest consists of packed solid grains, a wetting-phase fluid (including a free part and trapped/ adsorbed part), and air, the latter two together completely occupying the pore space. All constituents are permitted to be compressible, and thermal volumetric strain is considered for both the wetting fluid and solid grains. A combination of an isothermal condition, full wetting-phase saturation (no air), and no fluid entrapment reduce the problem to a limiting case where the material system configuration becomes identical to that for the general isothermal two-phase saturated poromechanics formulated above in section 2. We hereinafter focus on this limiting case in Abaqus.

### 3.1 Conservation laws

The wetting-phase fluid mass conservation law (i.e., continuity equation) is formulated over a Eulerian control volume containing a fixed amount of solid matter (i.e., the control volume is the solid material volume). This is equivalent to tracking the motion of solid particles. The fluid mass balance is sought between the total rate of change in liquid mass within the control volume and the liquid mass across the associated control surface. Adjusted to our nomenclatures and notations, it reads (here we specifically denote the $d/dt$ as $d^s/dt$ to highlight the reference frame)

$$\frac{1}{J}\frac{d^s}{dt}\left(J\rho_f\phi_f\right) + \nabla\cdot\left(\rho_f\phi_f\tilde{\boldsymbol{v}}\right) = 0 \tag{36}$$

The quasi-static force equilibrium in Abaqus is formulated directly in the weak form following the principle of virtual work (rate) over the same current control volume $V$. In our limiting case, it reads

$$\int_V \boldsymbol{\sigma}:\nabla^{(s)}\delta\boldsymbol{v}_s\, dV = \int_S \mathbf{t}\cdot\delta\boldsymbol{v}_s\, dS + \int_V \mathbf{f}_s\cdot\delta\boldsymbol{v}_s\, dV + \int_V \rho_f\phi_f\mathbf{g}\cdot\delta\boldsymbol{v}_s\, dV \tag{37}$$

where $\delta v_s$ is the virtual solid particle velocity, $\mathbf{t}$ is the traction on the current control surface $S$ and $\mathbf{f_s}$ is the body force excluding the part due to the liquid weight (i.e., solid weight, essentially $\rho_s(1-\phi_f)\mathbf{g}$, although this is not specifically written out in the manual).

It can be readily seen that the corresponding strong form of equation (37) is identical to equation (22) subjected to a standard Neumann boundary traction $\mathbf{t}$. However, under full saturation, Abaqus employs a simple effective stress principle for the porous solid, which reads in our limiting case as

$$\bar{\boldsymbol{\sigma}} = \boldsymbol{\sigma} + p_f\mathbf{1} \tag{38}$$

- 11 -

Substituting equation (38) into the strong form of equation (37) (same as (22)) and carrying out simple manipulations leads to the force balance equation as

$$\nabla \cdot \overline{\boldsymbol{\sigma}} - \nabla p_f + \left(\rho_f \phi_f + \rho_s \left(1 - \phi_f\right)\right)\mathbf{g} = \mathbf{0} \tag{39}$$

**3.2 Constitutive laws**

Abaqus implements the nonlinear Forchheimer's law and its linearized version - the Darcy's law (identical to equation (30)) - to describe the macroscopic fluid behavior. On the other hand, for solid grains, Abaqus adopts the concept of *effective strain*, which is defined as the total strain tensor minus/plus three isotropic parts resulting from volumetric changes to solid grains due to fluid pressure, thermal expansion/contraction, and fluid entrapment-induced void ratio variations, as well as a fourth saturation-driven moisture swelling anisotropic part. In our limiting case, the last three parts vanish and the effective strain, denoted as $\overline{\boldsymbol{\varepsilon}}$, takes the form of

$$\overline{\boldsymbol{\varepsilon}} = \boldsymbol{\varepsilon} + \frac{1}{3}\frac{p_f}{K_s}\mathbf{1} \tag{40}$$

where $\boldsymbol{\varepsilon}$ is the same strain tensor as in equation (32).

In Abaqus, it is this effective strain that is assumed to modify the effective stress (denoted as $\overline{\boldsymbol{\sigma}}$) for the solid. For a linearly elastic solid skeleton, that is

$$\overline{\boldsymbol{\sigma}} = \mathbb{C}^e : \overline{\boldsymbol{\varepsilon}} \tag{41}$$

where $\mathbb{C}^e$ is the same 4$^{th}$-order elastic stiffness tensor as in equation (31).

Finally, to account for material compressibility, Abaqus defines two sets of laws. Ignoring thermal expansion, the fluid compresses linearly with pore pressure, whereas solid grains compress linearly with effective mean stress and fluid pressure simultaneously. They read

$$d\rho_f = \frac{\rho_{f0}}{K_f} dp_f \tag{42}$$

$$d\rho_s = \frac{\rho_{s0}}{K_s}\left(-\frac{1}{3(1-\phi_{f0})}d\left(tr(\overline{\boldsymbol{\sigma}})\right) + dp_f\right) \tag{43}$$

Here, $K_f$ and $K_s$ are the bulk moduli of the wetting liquid and solid grains, respectively.

## 4. Equivalency

Here, we demonstrate the equivalency between the conservational frameworks of the general saturated two-phase poromechanics in section 2 and the limiting case of Abaqus soil mechanics in section 3. We first show that the general forms of mass conservation laws in the two problems are equivalent. To see this, we consider equation (9) and expand the first term on the *LHS* in equation (36) as

$$\frac{1}{J}\frac{d^s}{dt}\left(J\rho_f\phi_f\right) = \frac{1}{J}\left[J\frac{d^s}{dt}\left(\rho_f\phi_f\right) + \left(\rho_f\phi_f\right)\dot{J}\right] = \frac{d^s}{dt}\left(\rho_f\phi_f\right) + \left(\rho_f\phi_f\right)\nabla\cdot\boldsymbol{v}_s \tag{44}$$

Clearly, equation (36), upon substitution of equation (44), becomes identical to equation (17). Furthermore, their specific forms are also identical when both implement the Darcy's law equation (30).

On the other hand, the general forms of force balance law in the two problems appear different, as suggested by their respective non-body force terms on the *LHS* in equations (24) and (39). However, when assuming the solid skeleton as linearly elastic, both take the same specific forms. To demonstrate this, we substitute equations (40) and (41) into equation (39), yielding the following expression of its first two terms

$$\nabla\cdot\bar{\boldsymbol{\sigma}} - \nabla p_f = \nabla\cdot\left(\mathbb{C}^e:\boldsymbol{\varepsilon} + \frac{1}{3}\frac{p_f}{K_s}\mathbb{C}^e:\mathbf{1}\right) - \nabla p_f \tag{45}$$

Here, double tensor contraction between $\mathbb{C}^e$ and $\mathbf{1}$ leads the following reduction

$$\mathbb{C}^e:\mathbf{1} = K\mathbf{1}\otimes\mathbf{1}:\mathbf{1} + 2\mu\underbrace{\left(\mathbf{I} - \frac{1}{3}\mathbf{1}\otimes\mathbf{1}\right):\mathbf{1}}_{0} = K\mathbf{1}tr(\mathbf{1}) = 3K\mathbf{1} \tag{46}$$

Therefore, equation (45) now becomes

$$\nabla\cdot\bar{\boldsymbol{\sigma}} - \nabla p_f = \nabla\cdot\left(\mathbb{C}^e:\boldsymbol{\varepsilon}\right) + \nabla\cdot\left(\frac{K}{K_s}p_f\mathbf{1}\right) - \nabla p_f = \nabla\cdot\left(\mathbb{C}^e:\boldsymbol{\varepsilon}\right) - \left(1 - \frac{K}{K_s}\right)\nabla p_f \tag{47}$$

The equivalency is established by substituting equation (47) into equation (39), and equation (31) into equation (24).

To summarize, in the two poromechanical problems, the fluid sub-problem (with two orders of solid-to-fluid coupling) is always equivalent, and the solid sub-problem (with two orders of fluid-to-solid) problem is equivalent in the linearly poroelastic limit.

## 5. Relation to Biot Poroelasticity

### 5.1 Biot's Framework

The seminal theory of Biot poroelasticity (Biot, 1941) is widely adopted in analyzing coupled hydro-geomechanical problems. Two critical postulations underpin the theory to give rise to the fluid-solid full coupling. They are two phenomenological sets of linear constitutive laws initially conceived for a porous solid saturated with an incompressible fluid,

$$\theta = \frac{1}{3H} tr(\sigma) + \frac{1}{R} p_f = \alpha tr(\varepsilon) + \frac{1}{Q} p_f \tag{48}$$

$$\varepsilon = \mathbf{S} : \sigma + \frac{1}{H} p_f \mathbf{1} \tag{49}$$

where $H$ and $R$ are material constants albeit with unclear physical meanings, $\alpha$ is the Biot coefficient, $Q$ is the Biot modulus (denoted alternatively as $M$ in many studies), $\mathbf{S}$ is the elastic compliance tensor of the soil skeleton, $\sigma$ and $\varepsilon$ are same as before, $\theta$ is the increment of water content per unit volume of the porous solid. $\theta$ is rephrased by Rice & Cleary (1976) as the apparent volumetric fraction.

For convenience, equation (49) is reformulated as the widely known Biot effective stress law (again the sign convention is tension positive) by exploring relations amongst material constants,

$$\sigma = \mathbb{C}^e : \varepsilon - \alpha p_f \mathbf{1} = \sigma' - \alpha p_f \mathbf{1} \tag{50}$$

The mass balance law adopted by Biot takes the simplest Eulerian form when the fluid is compressible. It reads

$$\frac{\partial \theta}{\partial t} + \nabla \cdot \tilde{\underline{v}} = 0 \tag{51}$$

where $\tilde{\underline{v}}$ is the Darcy velocity written in our nomenclature. Biot originally used an isotropic permeability in Darcy's equation, and the term $\nabla \cdot \tilde{\underline{v}}$ simply becomes $k\nabla^2 p_f$ where $k$ is the isotropic permeability (see also text below equation (30)). Here, we do not expand this term. Substituting equation (48) into equation (51) recovers the form of mass balance given by Biot

$$\frac{1}{Q} \frac{\partial p_f}{\partial t} + \alpha \frac{\partial}{\partial t} tr(\varepsilon) + \nabla \cdot \tilde{\underline{v}} = 0 \tag{52}$$

The quasi-static force balance law given by Biot is obtained by letting the total stress $\sigma$ self-balance while ignoring body forces, that is, $\nabla \cdot \sigma = \mathbf{0}$, which, after substituting in equation (50), essentially reads

$$\nabla \cdot \sigma' - \alpha \nabla p_f = \mathbf{0} \tag{53}$$

Equations (52) and (53) are monolithically coupled. Specifically, the solid-to-fluid coupling is achieved by the term $\alpha \partial (tr(\varepsilon))/\partial t$ in equation (52), and the fluid-to-solid coupling by $-\alpha \nabla p_f$ in equation (53).

If the elastic porous solid skeleton is also isotropic, then four independent material properties are required to fully constrain the problem, two of which are standard elastic properties (such as shear modulus $\mu$ and Poisson's ratio $v$, or other pairs). Combinations of the remaining two parameters can be $H$ and $R$, or $\alpha$ and $Q$. The pore fluid introduces additional independent parameters, with the most notable one being the permeability tensor $\boldsymbol{k}$.

A notable subsequent development is given by Rice & Cleary (1976) who introduced the distinction between drained and undrained elastic behaviors and reformulated the problem using a new set of four independent parameters that are physically more interpretable. One common combination includes the Skempton coefficient $B$ that relates pore pressure changes to total mean stress in an undrained setting (Skempton, 1954), drained and undrained Poisson's ratios $v$, $v_u$, and the shear modulus $\mu$ which is not altered by pore fluids. However, the two assumptions expressed by equations (48) and (49) remain fundamentally unchanged. In particular, equation (50) is retained except for the replacement of $\alpha$ with new parameters,

$$\alpha = \frac{3(v_u - v)}{B(1+v_u)(1-2v)} \tag{54}$$

and equation (48) is recast as

$$\Delta m = m - m_0 = \rho_0 \frac{3(v_u - v)}{2\mu B(1+v)(1+v_u)} \left( tr(\boldsymbol{\sigma}) + \frac{3}{B} p_f \right) \tag{55}$$

where $\Delta m$ is the change in the mass of pore fluid per unit of current mixture volume.

The mass balance law also retains the simple Eulerian form of equation (51) except for the inclusion of pore fluid density,

$$\frac{\partial m}{\partial t} + \nabla \cdot (\rho_0 \tilde{\underline{v}}) = 0 \tag{56}$$

Substituting equation (55) into equation (56) and canceling $\rho_0$ recovers the alternative form of mass balance

$$\frac{9(v_u - v)}{2\mu B^2 (1+v)(1+v_u)} \frac{\partial p_f}{\partial t} + \frac{3(v_u - v)}{2\mu B(1+v)(1+v_u)} \frac{\partial}{\partial t} tr(\boldsymbol{\sigma}) + \nabla \cdot \tilde{\underline{v}} = 0 \tag{57}$$

The alternative form of force balance is the same as equation (53) except for the replacement of $\alpha$ according to equation (54).

Many subsequent studies followed, fundamentally resting on equations (48) and (50) and retaining Biot's conservation framework established by equations (52) and (53) or their alternative forms by Rice & Cleary (1976). The physical meaning of $\alpha$ has been established by Nur & Byerlee (1971) based on equation (49), and more rigorously from thermodynamics principles by Coussy (2004) and Borja (2006), see also section 2.3. However, how Biot's linearly poroelastic framework relates to the general poromechanical framework established above remains unclear.

**5.2 Reduction**

In this section, we show that Biot's linearly poroelastic framework is a reduced form of our general poromechanical framework under simplifying conditions. We start with the force balance law for the fluid-solid mixture. It can be readily seen that equation (53) is a reduced form of equation (24), in that the body force term is neglected. Consequently, the 2$^{nd}$-order fluid-to-solid coupling effect is not captured by equation (53), either.

Perhaps a more intriguing question pertains to the pore fluid itself, specifically, how equation (52) in Biot's framework relates to equation (17) in our framework. To investigate this, we expand the equivalent form of equation (17), which is equation (20), and divide both sides with the pore fluid density $\rho_f$, leading to

$$\frac{d^f \phi_f}{dt} + \frac{\phi_f}{\rho_f} \frac{d^f \rho_f}{dt} + \phi_f \nabla \cdot \mathbf{v}_f = 0 \tag{58}$$

Here, utilizing the definition of compressibility, we have

$$\frac{1}{\rho_f} \frac{d^f \rho_f}{dt} = \frac{1}{\rho_f} \frac{d^f \rho_f}{d^f p_f} \frac{d^f p_f}{dt} = C_f \frac{d^f p_f}{dt} = \frac{1}{K_f} \frac{d^f p_f}{dt} \tag{59}$$

where $C_f$ is the compressibility of the pore fluid and relates to $K_f$ as $C_f = 1/K_f$.

By exploring relations provided by equations (7) and (13), and substituting in equation (59), equation (58) can be further expanded in a mixed (fluid-Lagrangian, solid-Lagrangian, and Eulerian) reference frame as

$$\frac{d^s \phi_f}{dt} + \tilde{\mathbf{v}} \cdot \nabla \phi_f + \frac{\phi_f}{K_f} \frac{d^f p_f}{dt} + \phi_f \nabla \cdot \mathbf{v}_s + \phi_f \nabla \cdot \tilde{\mathbf{v}} = 0 \tag{60}$$

Collapsing the 2$^{nd}$ and 5$^{th}$ terms, and considering the relation given in equation (30), equation (60) can be trivially manipulated into

$$\frac{d^s \phi_f}{dt} + \frac{\phi_f}{K_f} \frac{d^f p_f}{dt} + \phi_f \nabla \cdot \mathbf{v}_s + \nabla \cdot \tilde{\underline{\mathbf{v}}} = 0 \tag{61}$$

This is another general form of the pore fluid mass balance under the three generic assumptions stated in section 2.2.

Within the range of infinitesimal deformation, we can utilize the fundamental incremental pore constitutive law equation (34), in which $d^s\varepsilon_{vol} \approx \nabla \cdot \boldsymbol{v}_s dt$. Considering this, and noticing equation (35), the rate of pore space increment reads

$$\frac{d^s \phi_f}{dt} = \frac{\alpha - \phi_f}{K_s} \frac{d^s p_f}{dt} + (\alpha - \phi_f) \nabla \cdot \boldsymbol{v}_s \tag{62}$$

Substituting equation (62) into equation (61) with further algebraic manipulation gives

$$\frac{\alpha - \phi_f}{K_s} \frac{d^s p_f}{dt} + \frac{\phi_f}{K_f} \frac{d^f p_f}{dt} + \alpha \nabla \cdot \boldsymbol{v}_s + \nabla \cdot \tilde{\boldsymbol{v}} = 0 \tag{63}$$

Interestingly, equation (63) naturally gives rise to the distinction between undrained and drained settings. Instead of being defined by flow boundary conditions, they can now be defined more intrinsically as

$$\frac{d}{dt} = \frac{d^s}{dt} \begin{cases} = \dfrac{d^f}{dt} (or, \boldsymbol{v}_f = \boldsymbol{v}_s) \leftarrow undrained \\ \neq \dfrac{d^f}{dt} (or, \boldsymbol{v}_f \neq \boldsymbol{v}_s) \leftarrow drained \end{cases} \tag{64}$$

In the undrained limit, equation (64) reduces to

$$\frac{1}{M} \frac{dp_f}{dt} + \alpha \nabla \cdot \boldsymbol{v}_s + \nabla \cdot \tilde{\boldsymbol{v}} = 0 \tag{65}$$

Here,

$$\frac{1}{M} = \frac{1}{Q} = \frac{\phi_f}{K_f} + \frac{\alpha - \phi_f}{K_s} \tag{66}$$

where $M$ (or $Q$) is the Biot modulus the same as before, and its expression is remarkably consistent with that provided in other studies assuming no-flow boundary conditions (e.g., Wang, 2000; Cheng, 2016).

Utilizing relations provided in equation (6), equation (65) takes the following Eulerian form

$$\frac{1}{M} \frac{\partial p_f}{\partial t} + \underbrace{\frac{1}{M} \boldsymbol{v}_s \cdot \nabla p_f}_{\substack{2^{nd}-order \\ s-to-f\ coupling}} + \underbrace{\overbrace{\alpha \nabla \cdot \boldsymbol{v}_s}^{=\alpha \frac{\partial}{\partial t} tr(\varepsilon)}}_{\substack{1^{st}-order \\ s-to-f \\ coupling}} + \underbrace{\nabla \cdot \tilde{\boldsymbol{v}}}_{=0} = 0 \tag{67}$$

In the more general drained setting, utilizing again equation (6), substituting equation (13), and considering the relation between the Darcy velocity and the relative velocity shown in equation (30), equation (63) can now be recast into its Eulerian form. Algebraic manipulations lead to

$$\frac{1}{M}\frac{\partial p_f}{\partial t} + \frac{1}{M}\boldsymbol{v}_s \cdot \nabla p_f + \underbrace{\frac{1}{K_f}\widetilde{\boldsymbol{v}} \cdot \nabla p_f}_{\substack{2nd-order \\ fluid\ term}} + \alpha \nabla \cdot \boldsymbol{v}_s + \underset{\neq 0}{\nabla \cdot \widetilde{\boldsymbol{v}}} = 0 \qquad (68)$$

It is worth noting again that specific forms shown by equations (67) and (68) are made possible by utilizing the tangential pore constitutive law equation (34), which assumes infinitesimal deformation and isotropy of the porous solid skeleton. If provided with such conditions, then contrasting these two equations with equation (52) reveals the relationship between our general law and Biot's law for pore fluid mass conservation. In the undrained limit, our equation (68) suggests the presence of an additional 2nd-order solid-to-fluid coupling term but a vanishing divergence of fluid flux (which means no normal fluid flux across a given domain boundary). Notice that the pressure gradient in the 2nd-order solid-to-fluid coupling term is linearly related to the Darcy velocity (see equation (30)), therefore it essentially scales linearly with $\boldsymbol{v}_s \cdot \widetilde{\boldsymbol{v}}$. This means this coupling effect vanishes only when $\boldsymbol{v}_s$ and $\widetilde{\boldsymbol{v}}$ are orthogonal to each other and is prevalent otherwise. In the general drained case, in addition to the presence of the 2nd-order solid-to-fluid coupling, there also appears a 2nd-order fluid term while the divergence of fluid flux prevails.

## 6. Summary

We introduced a general set of conservation laws governing a fully saturated fluid-solid mixture. The material system is isothermal without inter-phase mass exchange or external (fluid) mass sources. The conservation of mass of fluid hosted within a moving and deforming porous solid skeleton was derived from fundamental transport rules, and naturally gives rise to undrained and drained distinction. The quasi-static force balance law for the mixture was obtained from established studies. The two conservation laws are naturally fully coupled without prerequisites, with both 1st- and 2nd-order coupling terms describing fluid-solid interactions. This poromechanical framework subsumes the canonical Biot's framework of poroelasticity, which not only exhibits just the 1st-order monolithic coupling but also inherently rests on two postulated linear constitutive laws to enforce the coupling. In the end, the difference in the force balance laws is straightforward, in that ours includes an evolving mixture body force term that captures the 2nd-order fluid-to-solid coupling. However, how the two mass conservation laws relate is more subtle. In the same Eulerian reference frame and within an infinitesimal deformation range, we illustrated that both share the same 1st-order solid-to-fluid coupling for a given homogeneous and isotropic solid skeleton, but our framework captures the 2nd-order solid-to-fluid coupling in the undrained limit, and an additional 2nd-order fluid term in the drained limit. Finally, our framework is also equivalent to the limiting-case soil mechanics problem formulated in Abaqus with full saturation and without fluid trapping /adsorption or thermal expansion of constituents. Specifically, the mass conservation laws are equivalent in a general sense and independent from fluid constitutive behaviors; the force balance laws are equivalent in their specific forms when assuming a linearly elastic porous solid skeleton.

To fully characterize the poromechanical system, either Biot's or our framework requires four independent parameters, in addition to a standard hydraulic property (permeability). In the Biot framework, a common set is the Skempton coefficient $B$ (or the Biot coefficient $\alpha$), drained and undrained Poisson's ratios $\nu$, $\nu_u$, and the shear modulus $\mu$. Our framework uses more intrinsic material properties, including the bulk moduli of the fluid and solid grains, $K_f$, and $K_s$, as well as two independent parameters of the dry (i.e., drained) porous skeleton such as its bulk modulus $K$ and Poisson's ratio $\nu$.

## Acknowledgement

I thank William J. Curry for proofreading. No data were used in producing the manuscript.